\documentclass[aps,pra,twocolumn,amsmath,amssymb]{revtex4} 
\usepackage[english]{babel}
\usepackage{latexsym}
\usepackage{graphics}
\usepackage{subfigure}
\usepackage{epsfig}
\begin{document}

\title{Mean-field theory of Bose-Fermi mixtures in optical lattices}
\author{H. Fehrmann$^{1}$, M. A. Baranov$^{1}$, B. Damski$^{2}$, M. Lewenstein$^{1}$, and  L. Santos$^{1}$}  
\address{(1) Institut f\"ur Theoretische Physik, Universit\"at Hannover, D-30167 Hannover,Germany\\
(2)Instytut Fizyki,
  Uniwersytet Jagiello\'nski, PL-30-059 Krak\'ow, Poland} 
\begin{abstract}

We extend the results of [M. Lewenstein {\it et al.}, Phys. Rev. Lett. \textbf{92}, 050401 (2004)] 
and determine the phase diagram of
a mixture of ultracold bosons and polarized fermions placed 
in an optical lattice using mean field theory. We obtain the 
analytic form of the phase boundaries separating the composite fermion phases  
that involve pairing of fermions with one or more
bosons, or bosonic holes, 
from the bosonic superfluid coexisting with Fermi liquid.
We compare the results with numerical simulations and discuss their validity
and relevance for current experiments. We present a careful discussion of experimental 
requirements necessary to observe the composite fermions and investigate 
their properties. 
\end{abstract}
\maketitle

\section{Introduction}

During the last few years, the intensive quest for the 
Bardeen-Cooper-Schrieffer (BCS) transition  \cite{RevBCS} in 
ultracold trapped Fermi gases has stimulated the interest 
in mixtures of fermions and bosons, 
from now on called Fermi-Bose (FB) mixtures. 
In these systems, sympathetic cooling techniques can be efficiently 
used to reach temperatures well down into the regime of quantum degeneracy 
\cite{bfcooling}. Recently, the rich physics of FB mixtures 
has become itself   
one of the central topics of the physics of ultracold gases. 
Various phenomena occurring in FB mixtures have been  
analyzed, including the phase separation between bosons 
and fermions \cite{molmer,stoof,pu}, the existence of novel types of 
collective modes \cite{stoof,pu,capuzzi,liu}, the appearance of effective 
Fermi-Fermi interactions mediated by the bosons \cite{stoof,pu,albus,viverit}, 
the collapse of the Fermi cloud in the presence of attractive interactions 
between bosons and fermions \cite{collapse}, 
or the effects characteristic for the 1D FB mixtures \cite{das,cazalilla}. 

Recently, the possibility to load ultracold atomic gases in 
periodic potentials induced by laser standing waves (optical lattices) 
has attracted considerable attention, motivated partially by the 
resemblances of  these systems to the 
solid-state ones \cite{BECLatt}. The remarkable experimental 
advances on atomic trapping and cooling on one side, and the possibility of 
manipulation of the interatomic interactions via Feshbach resonances 
\cite{Feshbach} on the other, have opened the way towards a 
fascinating new physics, namely that of {\em strongly correlated 
ultracold atomic gases in optical lattices}. The most important 
result so far in this brand new research field 
has been provided by the achievement of the superfluid (SF) to 
Mott-insulator (MI) transition in bosonic gases \cite{jaksch,bloch}.

The experimental success achieved in bosonic gases, has 
triggered the interest in the physics of BF mixtures in optical 
lattices, which under appropriate conditions can be well described 
by the Bose-Fermi Hubbard model (BFH) model \cite{eisert,blatter}. 
A particularly interesting feature resulting from the BFH model
is the possibility to produce fermionic composites, formed by the 
pairing of a fermion and a boson (or a fermion and a bosonic hole) 
\cite{svistunov,lewen,kagan}). 
A Bose-Fermi lattice gas is somewhat similar to the Bose lattice gas
of Refs. \cite{jaksch,bloch}, although it exhibits 
a much more complex and  richer behavior at low temperatures, 
as we have recently shown in Ref.~\cite{lewen}.
 
In Ref.~\cite{lewen}, we have discussed the limit of strong 
atom-atom interactions (strong coupling regime) 
at low temperatures, and have predicted the existence of novel
quantum phases in lattices that involve the previously mentioned 
composite fermions, which for attractive (repulsive) interactions 
between fermions and bosons, are formed by fermion and bosons (bosonic holes).
The resulting composite fermions present a remarkably rich phase diagram, 
and may form, depending on the system parameters, a normal Fermi liquid, a
density wave, a superfluid liquid, or an insulator with fermionic domains.

In this paper we significantly extend the analysis of 
Ref.~\cite{lewen}. In Sec.~\ref{sec:BFH} we present a description of the model, while in 
Sec.~\ref{sec:compo} we remind the readers the main results of Ref.~\cite{lewen}. 
In Sec.~\ref{sec:MF},  we determine 
the phase diagram of the system for arbitrary values of the 
chemical potential, the Fermi-Bose coupling, and 
the tunneling (hopping) amplitudes using mean-field theory \cite{sachdev}. 
In this sense our results can be 
considered as a generalization of the seminal analysis of the 
Bose-Hubbard model by Fisher {\it et al.} \cite{fisher}. We obtain the 
analytic form of the phase boundaries separating the composite fermion 
phases from the phase consisting on a bosonic superfluid coexisting with a 
Fermi liquid. The method we use is a "traditional" mean field approach, which
has a limited domain of applications, but has an advantage of being 
technically relatively simple. In the case considered in this paper, it can be rigorously 
confirmed using more sophisticated methods of functional integrals within the 
 quantum field theoretical framework. In Sec.~\ref{sec:num} 
we compare analytical and numerical results,  
and discuss their validity and relevance for current experiments. 
Sec.~\ref{sec:exp} is devoted to a careful discussion of the necessary requirements 
to observe the composite fermions and investigate 
their properties in experiments. Finally, we present our conclusions 
in Sec.~\ref{sec:conclu}.

\section{Bose-Fermi Hubbard model}
\label{sec:BFH}

In this section we present a description of the model in question.
Let us consider a sample of ultracold bosonic and 
(polarized) fermionic atoms trapped in an optical lattice, e.g.
$^7$Li-$^6$Li or $^{87}$Rb-$^{40}$K. Due to the 
periodicity of the lattice potential, the single-atom states form energy bands,
as in solid-state systems. If the temperature is low enough and/or the
lattice potential wells are sufficiently deep, 
the atoms can be assumed
to occupy only the lowest energy band. Of course, for fermions this is
only possible if their number is strictly smaller than the number of
lattice sites (filling factor $ \rho_F\le 1$). To describe the system
under these conditions, we choose a particularly suitable set of single
particle states in the lowest energy band, the so-called Wannier states,
which are essentially localized at each lattice site. The system is then
described by the tight-binding BFH model 
(for a derivation from a microscopic model, see Ref.~\cite{eisert}), 
which is a generalization of the fermionic Hubbard model, 
extensively studied in condensed matter theory (c.f. \cite {Auerbach}):
\begin{eqnarray}
&&H_{\mathrm{BFH}}=-\sum_{\left\langle ij\right\rangle }
(J_B b_{i}^{\dagger }b_{j}+J_F f_{i}^{\dagger }f_{j}+ {\rm h.c.}) \nonumber \\
&&+\sum_{i}\left[
\frac{1}{2}Vn_{i}(n_{i}-1)+Un_{i}m_{i}-\mu_B n_{i}-\mu_F m_i\right] , 
\label{Hamiltonian} 
\end{eqnarray}
 where $b_{i}^{\dagger }$, $b_{j}$, $f_{i}^{\dagger }$, 
$f_{j}$ are the bosonic and fermionic
creation-annihilation operators,  $n_{i}=b_{i}^{\dagger
}b_{i}$, $m_{i}=f_{i}^{\dagger }f_{i}$, and $\mu_B$ and $\mu_F$ are  the bosonic
and fermionic chemical potentials, respectively. 
The BFH model describes: i) nearest neighbor boson (fermion) hopping, with an  
associated negative energy, $-J_{B}$ ($-J_{F}$); 
ii) on-site repulsive boson-boson interactions with
an associated energy $V$; iii) on-site boson-fermion interactions with an
associated energy $U$, which is positive (negative) for repulsive
(attractive) interactions \cite{footnote0}.

For large hopping ($\gg U,V$) the low temperature state of the system in 2D and 3D 
consists of a superfluid of bosons, which condenses in the zero momentum mode, 
while the fermions form a Fermi liquid. At exact zero temperature, 
the fermions would presumably form a superfluid state due to the $p$-wave
pairing induced by the attractive boson-mediated fermion-fermion interactions
\cite{viverit}. In this letter we are interested in determining 
the boundary of this hopping-dominated fluid phase. This boundary 
is of primary experimental relevance, since it sets the possible 
conditions for the observation of the predicted interaction-dominated 
quantum phases \cite{lewen}. In order to determine this boundary, 
we have to extend the analysis of Ref.~\cite{lewen} to the 
previously unexplored regime of $J_{B,F}\simeq U,V$.
To this aim we formulate the mean-field theory for bosons following 
the line of Refs. \cite{sachdev,fisher}. We shall first   
consider the case of a small fermionic hopping ($J_F\simeq 0$) 
and a fermionic filling factor $\rho_F$ not too close to $1/2$.  
If $\rho_F\simeq 1/2$ the mean-field solution becomes unreliable due to 
nesting effects and (in 2D) also due to van Hove singularities.  

\section{Composite fermions and strongly correlated quantum phases}
\label{sec:compo}

Let us first recall the results of Ref.~\cite{lewen}, which concern the limit of  interactions much stronger than hopping.  Denoting 
$\alpha =U/V$, one obtains that in the case of vanishing hopping 
for $\tilde{\mu}-[\tilde{\mu}]+s>\alpha
>\tilde{\mu}-[\tilde{\mu}]+s-1$, $s$ holes (or alternatively $-s$ bosons) 
form with a single fermion the corresponding composite fermion, 
which is annihilated by 
$
\tilde f_i=\sqrt{(\tilde n-s)!/\tilde n!}(b_i^\dag)^s f_i
$
(
$
\sqrt{\tilde n!/(\tilde n-s)!}(b_i)^{-s} f_i
$
).
The inclusion of a small tunneling ($\ll U,V$) as a perturbation leads 
to an effective Fermi Hubbard Hamiltonian:
\begin{equation}
H_{eff}=-J_{eff}\sum_{\left\langle ij\right\rangle }(\tilde f_{i}^{\dagger
}\tilde f_{j} +{\rm h.c.} )+K_{eff}\sum_{\left\langle ij\right\rangle }\tilde m_{i}
\tilde m_{j},
\label{EffHamiltonian}
\end{equation}
where $\tilde m_i=\tilde f_i^\dagger \tilde f_i$, $-J_{eff}$ is 
the negative energy associated with the 
nearest neighbor hopping of composite fermions, and $K_{eff}$ 
is the energy corresponding to the  
nearest neighbor composite fermion-fermion interactions, which 
may be repulsive ($>0$) or attractive ($<0$)\cite{lukin}. 
This effective model is equivalent to that of spinless
interacting fermions (c.f. \cite{sachdev,Shankar}), and, despite its 
simplicity, has a rich phase diagram. 


For $K_{eff}>0$, and $\rho _{F}\ll 1$ (or $1-\rho _{F}\ll 1$), 
the ground state of $H_{eff}$ corresponds to a Fermi
liquid (a metal), and is well described in the Bloch representation. 
In this limit, as discussed in Ref.~\cite{lewen}, the system is
equivalent to a weakly-interacting (even for large $K_{eff}$) 
Fermi gas of spinless fermions (for $\rho _{F}\ll 1$), or
holes (for $1-\rho _{F}\ll 1$). 
The weakly-interacting picture becomes inadequate for 
$\rho_{F}\rightarrow 1/2$, and for large $\Delta=K_{eff}/2J_{eff}$, 
where the effects of the
interactions between fermions become important, and one expects the
appearance of localized phases. A physical insight on the properties of this
regime can be obtained by using Gutzwiller Ansatz (GA) \cite{lewen}.
Defining $\Delta _{crit}=(1+m_{z}^{2})/(1-m_{z}^{2})$, 
with $m_{z}=2\rho _{F}-1$, one can obtain that 
for $\Delta <\Delta _{crit}$ the ground state  
is a Fermi liquid, while for $\Delta >\Delta_{crit}$ it is a 
density wave with a period of two lattice sites.
We expect that the employed GA formalism predicts the phase boundary 
$\Delta _{crit}$ accurately for $\rho
_{F}$ close to $1/2$. For the special case of half filling, $\rho _{F}=1/2$, 
the ground state is the so-called checkerboard state.
For $K_{eff}<0$, 
and when $|\Delta |\ll 1$, and $\rho _{F}$ is close to zero (one), 
a very good approach to the ground state is given by a BCS ansatz,
in which the composite fermions (holes) of opposite momentum build $p$-wave
Cooper pairs. The ground state becomes more complex for arbitrary $\rho _{F}$, 
and for $\Delta $ approaching $-1$ from above. The system becomes strongly
correlated, and the composite fermions in the SF phase may build not only
pairs, but also triples, quadruples etc. The situation becomes, however, 
 simpler when 
$\Delta <-1$, since the composite fermions group into domains, 
forming what one can call a ''domain'' insulator.


\section{"Simple man's" mean field theory}
\label{sec:MF}

In order to observe the strongly correlated phases involving the composite fermions it is useful, if not necessary,  to know where are their boundaries. Thus, this question has both fundamental and practical character. In this section we will answer this question using perhaps the simplest version of mean field theory. We stress, however, that we have 
confirmed the results of this section using more sophisticated methods of functional integrals within the 
 quantum field theoretical framework.

In the following we shall consider the case of vanishing temperature, 
$T\rightarrow 0$, but 
still $k_BT \gg J_F$, in such a way that we can safely assume $J_F=0$.
In order to analyze the relevant intermediate case $J_B\simeq U,V$, we 
employ a mean-field formalism \cite{sachdev,fisher,stoof2}. 
We shall analyze only homogeneous phases (it is possible to prove that 
within the mean-field limit supersolid phases are energetically unfavorable). 
We introduce the superfluid order parameter 
$\psi=\langle b_i \rangle=\langle b_i^\dag \rangle$, for every site $i$. 
Then, neglecting higher order in the fluctuations, we can substitute 
$b_j^\dag b_i=\psi (b_j^\dag +b_i)-\psi ^2$. In this way, we can model
the properties of the complete Hamiltonian $H_{BFH}$ by a sum of 
single-site Hamiltonians:
$H_{MF}=\sum_i (H_{0i}-\psi{\cal W}_i)$,
where 
$
H_{0i}=\frac{V}{2}n_i(n_i-1)+Un_im_i
-\mu_Fm_i-\mu_Bn_i+2dJ_B\psi^2$,
${\cal W}_i=2dJ_B \left (b_i+b_i^\dag\right )$,
and $d$ is the spatial dimension.
The mean-field Hamiltonian $H_{MF}$ contains the same on-site terms 
as those of $H_{BHM}$ and additional terms, 
which represent the influence of neighboring sites. 

\begin{figure}[ht] 
\begin{center}
\psfig{file=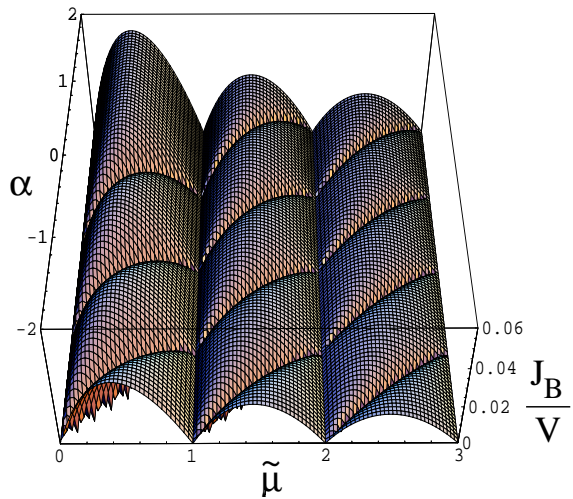,width=7.5cm}\\[0.2cm]
\end{center} 
\caption{Phase diagram in 2D as a function of $J_B/V$, the fermion-boson 
interactions $\alpha=U/V$, and the bosonic chemical potential $\tilde\mu=\mu_B/V$, 
for the case of $J_F=0$. The lobes denote the analytical phase boundaries 
calculated using our mean-field approach. For $\tilde n-1<\tilde \mu<\tilde n$, and  
$\bar{\mu}-\tilde n+s<\alpha<\bar{\mu}-\tilde n+s+1$,
the number of bosons, $n$, and the number of fermions, $m$, satisfy $n+sm=\tilde n$, 
and composites with one fermion and $s$ bosonic holes ($-s$ bosons) are formed.}
\label{fig:1}  
\end{figure}

As previously discussed, the ground-state of $H_{0i}$ consists of  
$n(m)=\tilde n-sm$ bosons per site, where $m$ is the number of fermions in the 
site, and $s$ depends on the particular region considered in the phase 
space. For a given fermionic filling factor $\rho_F$, the 
ground-state presents at every site $m=1$ ($m=0$) fermions with probability 
$\rho_F$ ($1-\rho_F$). Therefore the ground-state can be written as:
\begin{eqnarray}
|\phi_0\rangle\langle\phi_0|&=&(1-\rho_F)|n=\tilde n,m=0\rangle\langle
n=\tilde n,m=0| \nonumber \\ 
&+&\rho_F|n=\tilde n-s,m=1\rangle\langle n=\tilde n-s,m=1|.\nonumber
\end{eqnarray} 
The zeroth-order energy is of the form $E_0+2dJ_B\psi^2$, where  
$E_0=E_0(\tilde n,0)(1-\rho_F)+E_0(\tilde n-s,1)\rho_F$, with 
\begin{equation}
E_0(n,m)=\frac{V}{2}n(n-1)+Unm-\mu_Bn-\mu_Fm. 
\end{equation}
Due to the form of ${\cal W}_i$, the lowest order 
correction introduced by the tunneling occurs at second-order:
\begin{eqnarray}
E_2&=&\psi^2\sum_{n}\left\{
\frac{|\langle \tilde n,0|{\cal W}|n,0\rangle|^2}
{E_0(\tilde n,0)-E_0(n,0)}(1-\rho_F)
\right\delimiter 0
\nonumber \\ 
&&\left\delimiter 0
+
\frac{|\langle\tilde n-s,1|{\cal W}|n,1\rangle|^2}
{E_0(\tilde n-s,1)-E_0(n,1)}\rho_F
\right\}
\end{eqnarray}
where ${\cal W}=\sum_i{\cal W}_i$.
Following the Landau argument for second-order phase transitions, 
one can easily write \mbox{$E=E_0+2dJ_Br\psi^2+{\cal O}(\psi^4)$}, where 
\begin{eqnarray}
r&=&1+2dJ_B\left\{
\left (
\frac{\tilde n+1}{\epsilon(\tilde n,0)}-\frac{\tilde n}{\epsilon(\tilde n-1,0)}
\right )(1-\rho_F)
\right\delimiter 0
\nonumber \\
&&\left\delimiter 0
+\left (
\frac{\tilde n-s+1}{\epsilon(\tilde n-s,1)}-\frac{\tilde n-s}
{\epsilon(\tilde n-s-1,1)}
\right )\rho_F
\right \}
\end{eqnarray}
where $\epsilon(n,m)=\mu_B-Vn+Um$. 
If $r>0$ the system minimizes the energy by having $\psi=0$ (normal phase), 
whereas if $r<0$ a nonzero $\psi$ (superfluid) is energetically favorable. Therefore, 
the curve $r=0$ describes the boundaries between the phase with a superfluid 
bosonic gas, and the interaction-dominated phases.
In Fig.~\ref{fig:1} we have depicted the curves $r=0$ for different values of $\alpha$, 
and different regions of the $\mu_B-J_B$ phase space.

\section{Numerical results}
\label{sec:num}

The analytical results of the previous section have been compared with 
numerical calculations based in the GA approach. Technical details 
concerning the application of GA for fermions in 2D and 3D are discussed 
in App.~\ref{app:A}. The GA method allows a generalization of the 
analytical results presented above to the case of finite $J_F$, and for 
inhomogeneous phases. In our numerical approach, we first evaluate the 
Bloch functions for the lowest energy band, and properly combine them to 
obtain the corresponding Wannier functions, largely localized at 
each lattice site. 
Using the Wannier functions it is easy to obtain the coefficients 
$J_B$, $J_F$, 
$U$ and $V$ \cite{jaksch}. We employ the GA wavefunction:
$
|\phi\rangle=\prod_i \sum_{n_i=0}^{N_{max}} \sum_{m_i=0}^{1} 
f_{n_i,m_i} |n_i,m_i>,
$
where $n$ denotes the number of bosons and $m$ the number of fermions, 
and neglect fermionic anticommutation rules between operators at 
different sites (for 
a detailed discussion on this point, see App.~\ref{app:A}). 
We have checked that the value of the maximal bosonic occupation, $N_{max}$, 
does not affect the calculations. The coefficients $f_{n,m}$ must satisfy 
$\sum_{nm} |f_{n,m}|^2=1$. In the following, we consider only homogeneous phases, 
although the calculation can be straightforwardly 
extended to inhomogeneous phases. 
Hence, the ground-state can be found by minimizing the energy on a 
single cell, while using periodic boundary conditions. 
The energy is minimized by changing the values of the coefficients $f_{n,m}$ 
using a standard downhill technique.  
The ground-state is calculated for a given number of bosons and fermions, 
and for 
the case $U=0$, $V=0$ and a relative low lattice potential (large $J_B$). 
\begin{figure}[ht] 
\begin{center}
\psfig{file=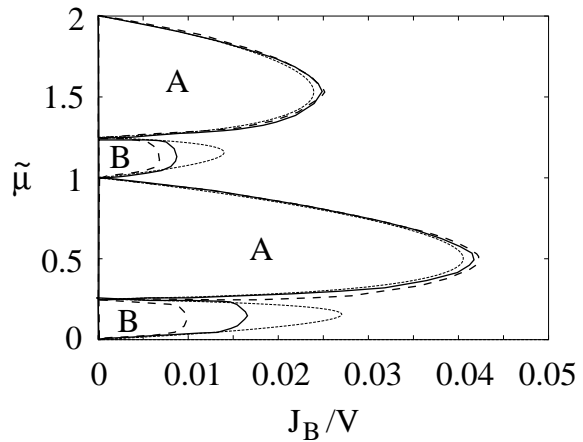,width=7.5cm}
\end{center} 
\caption{Phase diagram as a function of the hopping $J/V$, 
and the bosonic chemical potential $\tilde\mu=\mu_B/V$, for 
$\alpha=0.25$ and $\rho_f=0.25$. Thin solid lines: analytical results 
for $J_F=0$; Bold solid lines: numerical results for $J_F=0$; 
Dashed solid lines : numerical results for $J_F=J_B$.
Phases A are formed by a Mott-Insulator phase for the bosons and a Fermi liquid for the 
fermions. Phases B are characterized by the formation of fermionic composites 
with one fermion and one bosonic hole.}

\label{fig:2}  
\end{figure}

Starting from the chosen ground-state, we evolve in the phase space
by adiabatically varying the parameters of the system. The time evolution 
of the $f_{n,m}$ coefficients is obtained by employing the proper minimization 
$\frac{\partial }{ \partial f^*_{n,m} } \langle i\hbar \partial_t -H \rangle=0$. These equations  
determine the dynamics of the $f_{n,m}$ coefficients in the phase space, and constitute 
the basis of what has been called dynamic GA \cite{jaksch2}.
Since the total number of bosons and fermions is conserved during the time evolution, 
the chemical potential will be time dependent. This time dependence can be 
evaluated by calculating two initial ground-states 
$\psi_0 (t=0)$ $\psi_1 (t=0)$ 
with, respectively, a number of bosons $N_b$ and $N_b+ \delta N$, 
where $\delta N\ll N_b$. We evolve the parallel trajectories while  
reducing $J_B$. The chemical potential can be then approximated
as $\mu_b(t)\approx \left( \left< \psi_1 (t) | H(t)| \psi_1(t)\right> - 
\left< \psi_0 (t) | H(t)| \psi_0(t)\right> \right) /\delta N
$. By launching various trajectories, 
we can explore those regions with an incommensurate 
total number of bosons plus fermions. Consequently, the trajectories 
do not enter 
into the regions of the phase diagram in which 
commensurate phases are expected. Therefore 
the expected lobular gaps are opened. As shown in Fig.~\ref{fig:2}, 
our numerical and analytical results are in good agreement.
We must stress that the effects of the fermionic tunneling are not taken into 
account in the analytic calculations, but they can be easily included in our numerics. 
In Fig.~\ref{fig:2}, we also depict a case with finite fermionic tunneling ($J_B=J_F$). 
As expected, the lobes of the phases with composite fermions shrink
due to the larger mobility of the fermions. 

Finally we would like to comment about the validity of the mean-field approach presented in this 
paper. In general, the mean-field approach is exact at dimension $d=\infty$, and 
is expected to be reliable for $\tilde n\gg 1$, since the relative effects of fluctuations 
is then small. For the considered cases $\tilde n\sim 1$, in a general situation 
(not at the tips of the lobes) the upper critical dimension is $d_c=2$ \cite{fisher}, 
and therefore the mean-field approach is reliable. However, at the tips the transition  
belongs to a different universality class with $d_c=3$ \cite{fisher}, and hence 
our mean-field approach provides only a qualitative picture.

\section{Experimental requirements}
\label{sec:exp}

Let us finally consider in more detail the necessary requirements to observe 
the composite fermions and to study their properties. Let us use the notation 
of Ref.~\cite{lewen}, according to which Roman numbers denote how many particles 
form a composite, and a bar over the number indicates the composites fermion-bosonic hole(s). 
In the following discussion we shall assume $J_F\simeq J_B=J$.
We shall consider the regions of the phase diagram which are accessible at easiest:

\begin{itemize}
\item Region I : In this region boson-fermion interactions are weak, and 
 we deal with bare fermions. The hopping rate is of order $J$, i.e. 
relatively big in comparison to the effective interaction energy $K_{eff}$ 
which is of order $J^2/V$.

\item Region II : The boson-fermion interactions are attractive, and we 
deal with composites consisting of one fermion and one boson. 
Both the hopping rate $J_{eff}$, and the effective interaction energy 
$K_{eff}$ are of order $J^2/V$.

\item Region $\bar{\rm  II}$ : In this region the boson-fermion interactions 
are repulsive, and the composites consist of one fermion and one bosonic hole. 
As in Region II, the hopping rate  $J_{eff}$, and the effective interaction 
energy $K_{eff}$ are of order $J^2/V$.

\end{itemize}

Note, that in the regions III, IV, etc. the effective interactions $K_{eff}$ are 
of order $J^2/V$, but the effective hopping becomes of order $J^3/V^2$,  $J^4/V^3$, 
and so on. The 
larger the composites are obviously the less mobile, and for this reason the observation 
of their physics is more difficult than for regions I, II or $\bar{\rm  II}$.  
 
If we limit ourselves to  the Bose-Fermi mixtures in the regime $J<V,U$,
 there will be then the following temperature regimes:

\begin{enumerate}
\item  $T>|U|,V$ - This is the high temperature case. The system is a non-degenerated gas 
of fermions and bosons.

\item $T<|U|,V$ - The bosons enter the MI phase, and the fermions form composites (or remain 
bare fermions in the region I). Taking  into account the fact that the Fermi 
energy for the effective particles is provided by $J_{eff}$, we have to distinguish two cases:
\begin{enumerate}
\item When $T>J$, the composites constitute a non-degenerated Fermi gas in 
all considered regions I, II or $\bar{\rm  II}$ of the phase diagram.

\item For $J^2/V<T<J$, in the region I a degenerated gas is achieved, 
since $T<T_F$. Since $T>K_{eff}$, however,  this gas will behave as a practically  
ideal gas, since the interaction effects will be  masked by the thermal ones. Note, 
that in the other regions,  $T>T_F$, because $T_F\simeq
J_{eff}$. In another words, since the composites in the region II and  $\bar{\rm  II}$ 
have a large effective mass, we shall deal in this regime of temperatures with a 
practically  ideal non-degenerated gas of composites.

\item For $T<J^2/V$, in all regions I, II and $\bar{\rm  II}$ a 
degenerated interacting gas of composite fermions is formed, 
since $T<T_F$ and $T<K_{eff}$. 
\end{enumerate}
\end{enumerate}

In order to discuss the feasibility of these different temperature regimes, 
let us introduce the following notation:
Let $\sigma=V_0/E_{rec}$ be the 
optical lattice potential $V_0$ in units of the photon recoil energy $E_{rec}$, 
$a_z$ - the transversal oscillator length, and $a$ - the atomic scattering 
length (which, in principle, may be modified using Feshbach-resonance
techniques). Approximating the on-site potential by a harmonic oscillator, 
we obtain that the interaction energy entering the Hubbard model is of the form 
$$
V/E_{rec}=\sqrt{\frac{\sigma}{2\pi}}\frac{a}{2a_z}.
$$

According to the results of the present paper, one needs $J/V < 0.05$
in order to enter into the strongly correlated phases.
That means, in turn, that in order to reach the temperatures in which 
a degenerated gas of composite Fermions may be observed, i.e.  
$T<J^2/V$, one needs $T<2.5\times 10^{-3}\times\sqrt{\sigma/2\pi}\times(a/2a_z) 
T_{rec} \equiv T_0$. We recall that for typical experiments, for Lithium atoms 
$T_{rec}=2.11$ $\mu$K, whereas for Rubidium $T_{rec}=0.17$ $\mu$K. Obviously, lighter 
atoms (Li) are more favorable. 

Of course, the other parameters may be adjusted in various ways. 
In particular, apart from Feshbach technique, modifications of the 
transversal traps and lattice height can be used to control the atomic interactions. 
We have checked that, for instance, such manipulations in the case of 
$^6$Li-$^7$Li mixture, the value of $\alpha=U/V$ may be modified by a
 factor $4$. This can be easily seen, since
$$
\alpha=  \frac{2*A*(1+B)}{(1+\sqrt{B*G})(1*B*D)},$$
where $A=a_{BF}/a_{BB}$ is the ratio of scattering lengths,  $B=m_B/m_F$ is 
the ratio of the masses, $G=V_{0B}/V_{0F}$ is the ratio of optical 
lattice potentials, 
and $D=\omega_{zB}/\omega_{zF}$ is the ratio of transverse trapping 
frequencies for bosons and fermions, respectively. 

For instance for  $^6$Li-$^7$Li mixture $B=7/6$, and taking $G=0.3,D=0.3$, 
we get $\alpha=2a_{BF}/a_{BB}$, we gain a factor
of 2.
Assuming some concrete values of the parameters: $a_{BB}=500a_0$ (this value demands 
the employ of a Feshbach resonance), $a_{BF}=30a_0$
(with $a_0$ being the Bohr radius), we get  $\alpha=3/25$. 
If $\rho_F=0.5$ and $\omega_z=100$kHz, one obtains then for the region I, 
$U_{eff}=14.6{\rm nK}$, $U_{BB}=51 \mu{\rm K}$, $U_{BF}=6.1 \mu{\rm K}$, $J=21.5 {\rm nK}$.

The above estimates imply that	
the previously introduced temperature  regimes will be correspond to 
(1) $T>6$ $\mu$ K,
(2) $T < 6\ \mu$K,   
(2b) $T<21$nK
(2c) $T< 15$nK. 
From these estimations, and the current experimental state-of-the-art, we conclude that 
phase (2a), i.e. fermionic composites forming an ideal non-degenerated gas, should be 
relatively easy to achieve experimentally, since temperatures of the order of $1\mu$K are 
routinely obtained in current experiments. We would like to stress that such possibility alone 
could constitute a very interesting experiment by itself. The values to achieve degenerated 
phases, of the order of $10$nK, are more demanding experimentally, especially due to the difficulties 
to cool fermions imposed by Pauli-blocking. Nevertheless, this regime of temperatures 
is definitely not unrealistic, being at the border of what is nowadays achievable. 
The continuous developments in sympathetic cooling of fermions and bosons allow to foresee 
that such a regime could be achieved routinely in optical lattices in the nearest future. 

\section{Conclusions}
\label{sec:conclu}

In conclusion, we have completed the analysis of Ref.~\cite{lewen} by including the tunneling 
effects within the mean-field theory. This analysis is of crucial experimental importance 
since sets the conditions for the observability of the interesting phases 
with composite fermions. We have also developed a numerical method which is in good 
agreement with our analytics, and additionally provides information for those phases 
with finite fermionic hopping. Finally, we have analyzed the experimental requirements 
to observe the predicted effects. In particular, we have discussed different scenarios 
depending on the system temperature.

We would like to stress that in this paper we have always 
considered an homogeneous system (no additional external confinement). 
The analysis of finite temperature effects and inhomogeneities caused by 
the trap potential \cite{mura}  is 
interesting in itself, and has been recently  the subject of 
investigations in Ref. \cite{jens}. Particularly interesting is that a random on-site 
chemical potential in the presence of Fermi-Bose mixtures can lead to the achievement of 
various regimes of the physics of quantum fermionic 
disordered systems \cite{spinglass}.  

\acknowledgments
We thank  M. Cramer, J. Eisert, H.-U. Everts, M. Wilkens,  
and J. Zakrzewski for fruitful discussions.
We acknowledge support from the Deutsche
Forschungsgemeinschaft SFB 407 and SPP1116, the RTN Cold Quantum Gases, 
IST Program EQUIP, ESF PESC BEC2000+, and the Alexander von Humboldt 
Foundation.

\appendix
\section{Validity of simple man's Gutzwiller ansatz}
\label{app:A}

In this Appendix we discuss some technical details concerning 
the simple man's Gutzwiller ansatz for fermions, consisting in writing the
variational wave function as a product of on-site Fermi operators, 
and neglecting the anti-commutation relations between Fermi 
operators in different sites. 

Formally, our variational approach is equivalent to replacing the 
fermionic annihilation and creation operators $\tilde f_i,\tilde f_i^{\dag}$ 
by spin 1/2 operators $\sigma(i), \sigma^{\dag}(i)$. This approach can
 be justified using the exact Jordan-Wigner transformation in 2D and 3D 
(\cite{fradkin, eliezer,huerta}, see also \cite{tvselik}). In the case 
of 2D this 
transformation acquires the form
\begin{eqnarray}
S^+(j)&=&\tilde f^{\dag}_j\exp\left [ i\sum_{k\ne j}\arg(k,j)\tilde 
f_k^\dag 
\tilde f_k \right ],\\
S^-(j)&=&\exp\left[-i\sum_{k\ne j}\arg(k,j)\tilde f_k^\dag 
\tilde f_k\right ]\tilde f_j,\\
S_z(j)&=&\tilde f^{\dag}_j \tilde f_j-\frac{1}{2},  
\label{jw-traf}
\end{eqnarray}
where $\arg(k,j)$ is the angle between ${\bf k}-{\bf j}$ and an arbitrary 
space direction on the lattice, which we choose as $x$. 
The fermionic Hamiltonian of Eq. (\ref{EffHamiltonian})
becomes then
\begin{eqnarray}
H_{eff}&=& J_{eff}\sum_{\langle i,j\rangle}S^+(j)e^{iA(i,j)}S^-(i) + h.c.
\nonumber\\
&+& K_{eff}\sum_{\langle i,j\rangle}\left ( S_z(i)+\frac{1}{2}\right )\left ( S_z(j)+\frac{1}{2} \right ),
\label{hamil}
\end{eqnarray}
where the ``magnetic vector potential''
\begin{equation}
A(i,j)=\sum_{k\ne i,j}[\arg(k,i)-\arg(k,j)](S_z(k)+ 1/2).
\label{poten}
\end{equation}
The simple man's Gutzwiller approximation corresponds on this level 
to i) a variational ansatz for the ground state wave function 
in the form of product of on-site
spin states, and ii) substitution of  the ``magnetic potential'' by its 
average, which under assumption of 
mirror reflection symmetry with respect to the lattice axes is zero. 
Similar approach may be applied in 3D, although in that case the 
3D Jordan-Wigner transformation requires an extended Hilbert 
space and non-Abelian gauge transformations \cite{tvselik}.


\begin{references}

\bibitem{RevBCS} G. V. Shlyapnikov, Proc. XVIII Int. 
Conf. on Atomic Physics, Eds.: H. R. Sadeghpour, D. E. Pritchard, and E. J. Heller, 
(World Scientific Publishing, Singapore, 2002).

\bibitem{bfcooling} A. G. Truscott {\it et al.}, 
Science {\bf 291}, 2570 (2001); 
F. Schreck {\it et al.}, Phys. Rev. Lett. {\bf 87}, 
080403 (2001);  
Z. Hadzibabic {\it et al.}, Phys. Rev. Lett. {\bf 88}, 
160401 (2002); 
G. Modugno {\it et al.}, Phys. Rev. A {\bf 68}, 011601 (2003);
Z. Hadzibabic {\it et al.},  Phys. Rev. Lett. {\bf 91} 160401 (2003).; 


\bibitem{molmer} K. M\o lmer, Phys. Rev. Lett. 80, 1804-1807 (1998). 

\bibitem{stoof} M. J. Bijlsma, B. A. Heringa and H. T. C. Stoof, 
Phys. Rev. A {\bf 61}, 053601 (2000).

\bibitem{pu} H. Pu {\it et al.}, Phys. Rev. Lett. {\bf 88}, 070408 (2002).

\bibitem{capuzzi} P. Capuzzi and E. S. Hern\'andez, Phys. Rev. A {\bf 64}, 043607 (2001).

\bibitem{liu} X.-J. Liu, and H. Hu, Phys. Rev. A 67, 023613 (2003).

\bibitem{albus}  A. Albus {\it et al.}, Phys. Rev. A {\bf 65}, 053607 (2002).

\bibitem{viverit} L. Viverit and S. Giorgini, Phys. Rev. A {\bf 66}, 063604 (2002).

\bibitem{collapse} G. Modugno {\it et al.}, 
Science {\bf 297}, 2240 (2002).


\bibitem{das} K. K. Das, Phys. Rev. Lett. {\bf 90}, 170403 (2003).

\bibitem{cazalilla} M. A. Cazalilla and A. F. Ho, cond-mat/0303550.

\bibitem{BECLatt} B. P. Anderson and M. Kasevich, Science 282, 1686 (1998); 
O. M\"orsch {\it et al.}, Phys. Rev. Lett. 87, 140402 (2001); 
W. K. Hensinger {\it et al.}, Nature 412, 52 (2001); 
F. S. Cataliotti {\it et al.}, Science, 293 843 (2001).

\bibitem{Feshbach}  S. Inouye \textit{et al.}, Nature (London)
\textbf{392}, 151 (1998); S. L. Cornish \textit{et al.}, 
Phys. Rev. Lett.
\textbf{85}, 1795 (2000).

\bibitem{jaksch}  D. Jaksch \textit{et al.}, Phys. Rev. Lett. \textbf{81}, 3108
(1998).

\bibitem{bloch}  M. Greiner \textit{et al.}, Nature \textbf{415}, 39 (2002).

\bibitem{eisert} A. Albus, F. Illuminati and J. Eisert, cond-mat/0304223.

\bibitem{blatter} H. P. B\"uchler and G. Blatter, Phys. Rev. Lett. {\bf 91}, 130404 (2003).

\bibitem{svistunov}  A. B. Kuklov and B. V. Svistunov, Phys. Rev. Lett. {\bf 90}, 100401 (2003). 

\bibitem{lewen} M. Lewenstein, L. Santos, M. Baranov and H. Fehrmann, Phys. Rev. Lett. {\bf 92}, 050401 (2004). 

\bibitem{kagan} This phenomenon, related to the appearance of 
counterflow superfluidity in Ref. \cite{svistunov}, 
may occur also in the absence of the optical lattice,  
M. Yu. Kagan, D. V. Efremov, and A.V. Klaptsov, cond-mat/0209481.


\bibitem{sachdev}  S. Sachdev, \textit{Quantum Phase Transitions},
(Cambridge University Press, Cambridge, 1999).

\bibitem{fisher} M. P. A. Fisher, P. B. Weichman, G. Grinstein, and D. S. Fisher, 
Phys. Rev. B {\bf 40}, 546-570 (1989)

\bibitem{footnote0} For $U<0$, the single-band model requires $|U|$ 
smaller than the band gap.

\bibitem{Auerbach}  A. Auerbach, \textit{Interacting Electrons and
Quantum magnetism}, (Springer,New York, 1994). 

\bibitem{lukin}  L.-M. Duan, E. Demler, and M.D. Lukin, Phys. Rev. Lett. 
{\bf 91}, 090402 (2003).


\bibitem{Shankar}  R. Shankar,  Rev. Mod. Phys, \textbf{66}, 129 (1994).

\bibitem{stoof2} D. van Oosten, P. van der Straten, and H. T. C. Stoof, Phys. Rev. A 63, 053601 (2001). 

\bibitem{jaksch2} D. Jaksch {\it et al.}, Phys. Rev. Lett. {\bf 89}, 040402 (2002).

\bibitem{mura}  G.G. Batrouni {\it et al.}, Phys. Rev. Lett. {\bf 89},
117203 (2002). 

\bibitem{jens} M. Cramer, J. Eisert,  and F. Illuminati, cond-mat/0310705.

\bibitem{spinglass} A. Sanpera {\it et al.}, cond-mat/0402375.

\bibitem{fradkin} E. Fradkin, Phys. Rev. Lett. {\bf 63}, 322 (1989).

\bibitem{eliezer} D. Eliezer and G. W. Semenoff, Phys. Lett. B {\bf 286}, 
118 (1992).

\bibitem{huerta}  L. Huerta and J. Zanelli, 
Phys. Rev. Lett. {\bf 71}, 3622 (1993).

\bibitem{tvselik} A. M. Tvselik, 
{\it ``Quantum field theory in condensed matter physics''}, 
Cambridge University Press, 1995.


\end{references}
\end{document}